\documentclass{article}
\usepackage{fullpage}
\usepackage{url}
\usepackage[hidelinks]{hyperref}
\usepackage[utf8]{inputenc}
\usepackage[small]{caption}
\usepackage{graphicx}
\usepackage{amsmath}
\usepackage{amsthm}
\usepackage{booktabs}
\usepackage{authblk}

\usepackage{color}
\usepackage{amsmath,amssymb,amsthm}
\usepackage{amsmath}
\usepackage[ruled,vlined,linesnumbered]{algorithm2e}

\usepackage{enumitem}

\newtheorem{theorem}{Theorem}
\newtheorem{lemma}{Lemma}

\newcommand{\remove}[1]{}

\newcommand{\dpj}[1]{{#1}}
\newcommand{\dk}[1]{{#1}}

\newcommand{\cO}{O}%{{\cal O}}

\newcommand{\polylog}{{\rm \ polylog\ }}

\newcommand{\cN}{{\mathcal{N}}}

\newcommand{\cT}{{\mathcal{T}}}

\newcommand{\cR}{{\mathcal{R}}}

\newcommand{\dispereps}{{\xi}}
\newcommand{\secondthresh}{{\varphi}}
\newcommand{\threshdiff}{{\epsilon}}
\newcommand{\ops}[1]{\lambda(#1)}

\begin{document}
%\title{Light Agents Searching for Hot Information}
\title{Efficient Algorithm for Deterministic Search of Hot Elements\thanks{D. Pajak was supported by the National Science Centre, Poland—Grant Number 2019/33/B/ST6/02988.}}

	\author[1]{Dariusz R. Kowalski}
	\author[2]{Dominik Pajak}
	\affil[1]{School of Computer and Cyber Sciences, Augusta University, USA, dkowalski@augusta.edu}
	\affil[2]{Wrocław University of Science and Technology, Poland, dominik.pajak@pwr.edu.pl}
\date{}
\maketitle

\begin{abstract}

%Agent-based crawlers are commonly used in network maintenance and information gathering. In order not to disturb the main functionality of the system, whether acting at nodes or being in transit, they need to operate online, perform a single operation fast and use small memory.
%They should also be preferably deterministic, as crawling agents have limited capabilities of generating a large number of purely random bits. We consider a system in which an agent receives an update, typically an insertion or deletion, of some information
%upon visiting a node. On request, the agent needs to output hot information, i.e., with the net occurrence above certain frequency threshold. A desired time and memory complexity of such agent should be poly-logarithmic in the number of visited nodes and inversely proportional to the frequency threshold. Ours is the first such agent with rigorous analysis and a complementary almost-matching lower~bound. 
%\vspace*{-3ex}
When facing a very large stream of data, it is often desirable to extract most important statistics 
online in a short time and using small memory.
%%using the smallest possible memory. 
For example, one may want to quickly find the most influential users generating posts online or check if the stream contains many identical elements. In this paper, we study streams containing insertions and deletions of elements from a possibly large set $N$ of size $|N| = n$, that are being processed by online deterministic algorithms. At any point in the stream the algorithm may be queried to output
elements of certain frequency in the already processed stream. More precisely,
the most frequent elements in the stream so far. 
The output is considered correct if 
the returned elements 
it contains all elements with frequency greater than a given parameter $\secondthresh$ and no element with frequency smaller than $\secondthresh-\threshdiff$. We present an efficient online deterministic algorithm for solving this problem using $O(\min(n, \frac{\polylog(n)}{\threshdiff}))$ memory and $O(\polylog(n))$ time per processing and outputting an element.
%operation. 
It is the first such algorithm as the 
previous algorithms were either randomized, or processed elements in substantially larger time $\Omega(\min(n, \frac{\log n}{\threshdiff}))$, or handled only insertions and required two passes over the stream (i.e., were not truly online). 
Our solution is almost-optimally scalable (with only a polylogarithmic overhead) and does not require randomness or scanning twice through the stream.
We complement the algorithm analysis with a lower bound $\Omega(\min(n, \frac{1}{\threshdiff}))$ on required memory. 
\end{abstract}

 \begin{table*}[t]
\small
	\centering
	\begin{tabular}{llllll}
		\toprule
		Type of algorithm & Time per item & Memory & 
		%Supported 
		Operations & Reference\\\midrule
		Deterministic -- two passes 
		& 
		$O(\frac{1}{\secondthresh}\log (\secondthresh n))$ $^\star$
		%$O(\log \frac{1}{\secondthresh})$ amortized 
		& $O(\frac{1}{\threshdiff})$ & insert only &  \cite{MisraG82}\\\midrule
		Deterministic 
		& 
		$O(\frac{1}{\secondthresh}\log (\secondthresh n))$ $^\dagger$
		%$O(\log (n\secondthresh))$ amortized 
		& $O(\frac{\log (n / \threshdiff)}{\threshdiff})$ & insert only  &  \cite{MankuM02}\\\midrule
		Randomized LV & $O(1)$ expected & $O(\frac{1}{\threshdiff})$ &  insert only & \cite{DemaineLM02}\\\midrule
		Randomized MC -- approx & $O(\log \frac{1}{\sigma} )$ & $O(\frac{\log n}{\secondthresh \varepsilon^2} )$ &  insert only & \cite{CharikarCF02}\\\midrule
		Randomized MC -- approx & $O(\log n \log \frac{1}{\secondthresh \sigma} )$ & $O\left(\frac{\log n \log \frac{1}{\secondthresh\sigma}}{\varepsilon} \right)$ & insert \& delete &  {\cite{cormode2005s}}\\\midrule
		\dpj{Deterministic} & \dpj{$O\left(\frac{\log_{1/(\secondthresh\threshdiff)}n \log n}{\secondthresh\threshdiff}\right)$} & \dpj{$O\left(\frac{\log^3 n}{\secondthresh^2\threshdiff^2}\right)$} & \dpj{insert \& delete}  & \dpj{{\cite{GangulyM07}}}\\\midrule
		Deterministic & $O(\polylog n)$ & $O(\frac{\log^3 n}{\varepsilon} )$ & insert \& delete &  this paper\\\bottomrule
	\end{tabular}
	\caption{\label{table-results}Performance of best algorithms finding hot elements. LV and MC denote Las Vegas and Monte Carlo solutions, resp., while approx mean only approximated solutions. 
	$^*$ shows additionally that an {\em amortized} processing time is $O(\log \frac{1}{\secondthresh})$, while in $^\dagger$ the amortized time is $O(\log (n\secondthresh))$, as discussed in~\protect\cite{cormode2005s}.
	Amortized time denotes the total time for the whole stream processing divided by the number of processed~operations, and is a weaker measure that worst-case complexity considered in this work.  
	In the results of the existing papers a notation $k$ was sometimes used that corresponds to $\secondthresh = 1/(k+1)$. Notation $\sigma$ in the existing papers denotes the probability of failure. The result of~\protect\cite{MisraG82} and \protect\cite{DemaineLM02} consider the special case, where $\threshdiff = \secondthresh$.}
	\end{table*}

\section{The Model and the Problem}

Finding elements occurring above certain frequency $\secondthresh$, so called hot items, is one of the fundamental tools in mining online streams and histogram study,
c.f.,~\cite{IoannidisC93,IoannidisP95}.
It can also be applied in data warehousing, data mining and information retrieval, decision support systems, databases, caching, load balancing, network management, anomaly detection, and many others, c.f.,~\cite{DemaineLM02,FangSGMU98,GibbonsM99,KarpSP03}. 
\remove{%%%%%%%%%%%%%%%  START REMOVE  %%%%%
In all these areas the universe of elements/items $N$ is often very large (too large to store or process all of them), while there is still a demand for fast data processing as well as fast and most accurate decisions \dk{done by processing agents}.
In other words, the agents
%algorithms 
should be scalable with respect to $\log\; n$ rather than $n$, where $n=|N|$, and wrt other problem parameters; in particular, their memory and time performance should be proportional to a polynomial in $\log\; n$ and the inverse of the sought item frequency 
$\secondthresh$.

%In the core of online mining of streams are algorithms finding elements occurring with at least a certain frequency.
%This task is particularly challenging if the universe $N$ from which the stream items are drawn is very large. 
Universe $N$ could contain various types of elements: simple atomic elements, relationships (graph or hypergraph edges), or even more complex items. 
}%%%%%%%%%%  END  REMOVE  %%%%%%%%%%%%%

We consider a stream of operations, also called transactions, involving elements in the universe $N$ of size $n$. We do not limit the distribution of elements in the stream -- they could be created arbitrarily, even by an online adversary, which aims at ``fooling'' the 
agent
%algorithm 
processing the stream aiming to find hot elements. Each operation involves a single element $x\in N$ and could be either an insertion or deletion of this element (to/from some large data repository). 
Observe that if a large $\Theta(n)$ space is allowed at an agent, then a simple heap data structure could process each insertion or deletion operation in $O(\log\; n)$ time, and find the hot items in $O(\frac{1}{\epsilon} \log\; n)$ time, for any stream, c.f.,
%in the worst case 
%Aho et al.~
\cite{Aho1983}. 
However, in case of large universe $N$ (Big Data), such a solution is not practical.
Therefore, for more than 20 years the research in this area focused on finding a summary data structure, of sublinear (in $n$) size and processing/enlisting time.

\remove{%%%%%%%%%  START  REMOVE  %%%%%%%%
In this work we follow this line of research and address the following fundamental question:

\begin{quote}
Is there an efficient (scalable) deterministic online algorithm processing any stream of operations and enlisting hot elements 
%of frequency at least $\secondthresh$ 
on request? 
\end{quote}
}%%%%%%  END  REMOVE  %%%%%%%%%%%%%%%

There is, however, a subtle twist -- could sublinear algorithms return all and only hot elements?
%Cormode and Muthukrishnan~
\cite{cormode2005s} showed that enlisting {\em all and only} hot elements is impossible with sublinear memory $o(n)$.
(They were inspired by a seminal paper 
%by Alon et al. 
\cite{AlonMS96} proving that estimating highest frequency is impossible in sublinear memory $o(n)$.)
This also applies to randomized algorithms: any algorithm which guarantees outputting all and only hot elements with probability at least $1 -\sigma$, for some constant~$\sigma$,
must also use $\Omega(n)$ memory. 
This generalization uses a related result on the Index problem in communication complexity, c.f., 
%Kushilevitz and Nisan~
\cite{KN97}.
This argument implies that, if we are to use less than $\Theta(n)$ memory, then
we must sometimes output items with frequency smaller than~$\secondthresh$.
%which are not hot. 
%since we will endeavor to include every hot item in the output. 
%In our guarantees, we will instead
Therefore, the main challenge is:
%a desired and feasible 
%
%\begin{quote}

\vspace*{1mm}
\noindent
{\em To design an efficient (light) deterministic online 
%agent
\dpj{algorithm (agent)}
processing any stream of operations and, upon request, listing all elements of frequency at least $\secondthresh$ and no element~of~frequency~at~most~$\secondthresh - \threshdiff$.}
%of frequency at least $\secondthresh$ 
%on request.
%\end{quote}

\vspace*{1mm}
Light, or well-scalable, agent means that it should process an operation or output an element in time at most polylogarithmic in $n$ (i.e., $\log^c n$ for some constant $c>0$), as $n=|N|$ could be very large, while using memory linear in $1/\threshdiff$ and polylogarithmic in $n$.
%The solution should be 
\remove{%%%%%  START  REMOVE  %%%%%%%%
guarantee is to output 
all hot items (i.e., with frequency at least $\secondthresh$) and no
items which are far from being hot --
%will be output. 
%(with arbitrary probability) 
that is, no items 
%which has
with
frequency less than 
%$1/(k+1)-\epsilon$ 
$\secondthresh - \threshdiff$, 
%will be output, 
for some user-specified parameter $\epsilon$.
}%%%%%%%  END  REMOVE  %%%%%%%%%%%%
%
%Is there an efficient scalable algorithm processing streamed items online and enlisting elements of frequency at least $\secondthresh$ on request? Here by {\em scalable} we understand processing time being proportional to a {\em polylogarithm} of the universe size $n=|N|$, and with memory and time to enlist all requested elements being $\polylog n$ multiplied by the inverse of the requested frequency.
%This question has been answered in affirmative in a few cases:
Light 
%agents
\dpj{algorithms}
have already been designed in some cases:

%\vspace*{-1ex}
\begin{itemize}[leftmargin=0.8em,itemsep=0ex]
%labelwidth=-3em]
\item
When randomness is allowed, c.f.,~\cite{cormode2005s}; however, false positives and false negatives are possible; also, it is not known if the result hold against adversarial creation of the stream (i.e., if the adversary decides on consecutive elements in the stream online, seeing the past choices of the algorithm)
\item
When second pre-processing or processing in larger batches (so called window-based) is allowed, c.f.,~\cite{MisraG82,lin2005mining}
%on the stream; 
however, such agents are not 
%strictly~speaking
pure
online.
%\item
%When larger processing time proportional to $n$
\end{itemize}

%\vspace*{-1ex}
\paragraph{Our results.}
In this work (Section~\ref{sec:algo}) we design a deterministic 
\dpj{algorithm}
%agent, represented by specific data structures and algorithms, 
that overcomes all of the abovementioned obstacles: it is fully online (it does not go backwards or look ahead when processing current stream location), it does not use any random bits, and it works for arbitrary streams even created by an online adaptive adversary. It handles both insertions and deletions. Finally, it is also light and well-scalable (as we analyze formally in Section~\ref{s:analysis}), in the sense that it uses only polylogarithmic time per operation and returning a hot element while using only $O(\frac{\log^3\; n}{\threshdiff})$ local memory. This memory space is close to optimal, as we show in the proof of a lower bound in Section~\ref{sec:lower}. Table~\ref{table-results} compares performance of our agent with most relevant previous work. Finally, we discuss possible extensions (including multi-agent parallelization) and open problems in Section~\ref{sec:conclusions}. 
%and has close to optimal space complexity.

\paragraph{Model and problem.}
Consider an incoming very long stream of operations on elements of a very large universe $N$ of size $n=|N|$, where each operation is of a form $\mathsf{op}(x)$, for $\mathsf{op} \in \{\mathsf{insert}, \mathsf{delete}\}$ and $x \in N$. We assume that each operation contains a $O(\log\; n)$-bit identifier of an element $x$ involved; we will be using ``element'' and ``item'' interchangeably throughout this paper. 
%Items, in particular an item could be an edge of a graph (or edge with multiplicity in case of weighted graphs).  
The {\em net occurrence} of an item $x \in N$ 
in step
%at time 
$t$, denoted by $n_t(x)$, is the number of insertions of $x$ minus the number of deletions of $x$ in the first $t$ operations of the stream. The {\em frequency} of element $x$ in step~$t$ is denoted by $f_t(x) = n_t(x) / \dpj{t}$.
%\sum_{y \in N} n_t(y)$. 
The stream satisfies Basic Integrity Constraint, as defined in~\cite{cormode2005s}: in each round a frequency of any element is non-negative, in particular, the number of deletions never exceeds the number of insertions of an element.
%\footnote{Optional version -- instead of basic integrity constraint, our algorithm could work under assumption that if the frequency is at most $\delta$ we do not have to execute this operation in a round.}

%Notation: $n$ number of different items, $t$ the number of current round, $m_t,M_t$ the number of edges (the sum of their multiplicities, resp.) at the end of step $t$

%``Machine word'': a unit of memory/space that is sufficient to store round number, and parameters $n$, $m_t$, $M_t$

The problem of finding hot elements (also called frequent elements) is parametrized by $1 > \secondthresh > \threshdiff > 0$; we denote it by finding $(\secondthresh,\;\threshdiff)$-hot elements. The objective is to design an agent, consisting of data structures and algorithms, capable of processing the operations of the stream online in sequence (without the possibility of returning to already processed operations). At any point, 
%the data structure 
upon external request the agent should be able to return a set of frequent elements. We will say that the output of the algorithm is correct if the returned set contains $O(1/\secondthresh)$ elements, including all elements with frequency at least $\secondthresh$ and no element with frequency smaller than or equal to $\secondthresh - \threshdiff$. 

Performance of an agent is measured in terms of time to process a single operation from the stream, time to output all the $(\secondthresh,\;\threshdiff)$-hot elements, and the total local memory used. In the measurement, the atomic operation concerns so called {\em Machine word}: a unit of memory that is sufficient to store a single element, step number and all problem parameters; every basic operation on machine words, e.g., arithmetics, is accounted as $1$ in time complexity. 
We aim at time and memory efficient agents, i.e., performing each operation (or outputting an element) in time $O(\!\polylog n)$ and using $O(\frac{1}{\threshdiff}\polylog n)$ memory units (each storing a machine word).

%We want to find a data structure for this problem, using the smallest possible amount of memory, where each unit of memory is a register sufficiently large to store round number, and parameters $n$, $m_t$, $M_t$.

%-----

%There are three important characteristics to consider: the space used, the time to update the data structure following each transaction (the update time), and the time to produce the hot items (the query time).

\paragraph{Additional notation.} By $t$ 
%we will be denoting 
denote
the current step number (i.e., the number of operations of the stream that have already been handled by the algorithm). We will also use notation~$\gamma = \frac{\threshdiff}{6}$.\footnote{%
\dk{Auxiliary parameters in this work are chosen for convenience of mathematical analysis in our general streaming model, without harming asymptotic performance. Further optimization of constants could be possible through more detail case study and/or specific experimental optimization for selected datasets.}}

\subsection{Previous and Related Work}

\paragraph{Handling insertions and deletions.}
Cormode and Muthukrishnan
\cite{cormode2005s} proposed randomized online algorithm
with memory $O(k\log k \log n)$, processing each operation in time $O(\log k \log m)$, and outputting hot elements in time $O(k\log k \log m)$, where $k=\frac{1}{\secondthresh}-1$. The algorithms return no items with frequencies less than $\frac{1}{k+1}-\threshdiff$ with some user-specified probability. 
%\cite{GilbertKMS02b,CormodeM05b}
%studied the problem of approximating the frequencies of hot elements
%in the presence of insertions and deletions.
%\dk{??? Jak to sie ma do nas ???}
A deterministic summary structure for data streams in~\cite{GangulyM07} finds the most frequent elements however it requires space $O(\log^n / (\varphi^2 \varepsilon^2))$. In \cite{CormodeM05b}, the objective is to return an approximate frequency of any element using memory $\widetilde{O}(1/\varepsilon)$ and time $\widetilde{O}(1)$, but the error was proportional to $\varepsilon$ times the total frequency of all the elements. In~\cite{GilbertKMS02b} the considered problem is to return approximate quantiles of the data -- the solution uses space $O(1/\varepsilon^2)$.

Earlier results include the problem of histogram maintenance, which involves finding a piecewise-constant approximation of data distribution. The optimal histogram is close to the data vector in terms of $\ell_1$ or $\ell_2$ norms hence it approximates all the data points, whereas in the problem of hot elements the objective is to approximate the frequencies of only the most frequent elements.
Gibbons et al. 
\cite{GibbonsMP97} 
%[1997] 
were the first who considered insertions and deletions in the context of maintaining various histograms, however their methods need periodical access to large portion of the data in the presence of deletes.
Gibbons and Matias~
\cite{GibbonsM98,GibbonsM99} analyzed mainly insertion operations, but also performed experimental study in the presence of deletions.
Gilbert et al.  
\cite{GilbertGIKMS02a} designed and analyzed algorithms for maintaining
histograms with guaranteed accuracy and small space.
%\dk{??? Rezultaty ???}

\iffalse
\dk{Ten paragraf do zmiany:}
The methods in this article can yield algorithms for maintaining hot items,
but the methods are rather sophisticated and use powerful range summable
random variables, resulting in $k^{\log O(1)} n$ space and time algorithms where the
O(1) term is quite large. We draw some inspiration from the methods in this
article—we will use ideas similar to the “sketching” developed in Gilbert et al.
[2002a] \cite{GilbertGIKMS02a}, but our overall methods are much simpler and more efficient. Finally,
recent work in maintaining quantiles [Gilbert et al. 2002b] \cite{GilbertKMS02b} is similar to ours
since it keeps the sum of items in random subsets. However, our result is, of
necessity, more involved, involving a random group generation phase based on
group testing, which was not needed in [Gilbert et al. 2002b] \cite{GilbertKMS02b}. Also, once such
groups are generated, we maintain sums of deterministic sets (in contrast to
the random sets as in Gilbert et al. [2002b] \cite{GilbertKMS02b}), given again by error correcting
codes. Finally, our algorithm is more efficient than the $\Omega(k^2 log^2 m)$ space and
time algorithms given in Gilbert et al. [2002b] \cite{GilbertKMS02b}.
\fi

\paragraph{Insertion-only streams.}
For streams with only insertions, in a special case without a lower bound on the frequency of returned elements (which is equivalent to $\threshdiff = \secondthresh$), \cite{MisraG82} designed a deterministic algorithm with processing and enlisting time $O(k\log k)$ and memory $O(k)$ (which in this case equals to $O(1/\threshdiff)$). Their algorithm however is not fully online, as it requires a second pass on the stream. In the same model, \cite{DemaineLM02} proposed a single-pass randomized algorithm for finding frequent elements using $O(k)$ memory and $O(1)$ expected time for processing one item.

In the more general case, \cite{MankuM02} proposed a deterministic algorithm Lossy Counting that processes finds hot elements using $O( \log(n/\threshdiff)/\threshdiff)$ memory. \cite{CharikarCF02} proposed a randomized algorithm using $O(k \log n/\threshdiff^2)$ memory and and $O(\log(1/\sigma))$ time per operation that succeeds with probability at least $1-\sigma$.

\paragraph{Recent results} \hspace*{-1em}
%More recently, results 
on finding frequent items in streams include models where items that are more recent in the stream have higher weight \cite{wu2017novel,cafaro2017frequency}, as well as many applications, e.g., finding frequent elements in two-dimensional data streams~\cite{lahiri2016identifying,epicoco2018fast}. 
\newcommand{\cC}{{\mathcal{C}}}
\newcommand{\cG}{{\mathcal{G}}}
\newcommand{\cH}{{\mathcal{H}}}
\newcommand{\op}{\textsf{op}}

\section{Algorithm}
\label{sec:algo}

%%%%%%%%%%%%%%%%  END  REMOVE  %%%%%%%%%%%%%%%%%%%

\subsection{Data Structures with Operations}

%\paragraph{Data structures.}
Our 
%algorithm 
agent
maintains two data structures: a \emph{disperser-based} structure of \emph{group counters} and a 
%\emph{heap-based} 
structure of \emph{individual entries} built on the top of \emph{balanced Binary Search Trees}. We define them as follows.

\paragraph{{\em Group counters.}} To build this structure, we use a 
is an {\em $(\ell,d,\dispereps)$-disperser graph $G = (V,W,E)$ with entropy loss $\delta$} for parameters:
$|V| = n$, 
%$|W| \ge \frac{c}{\gamma} \cdot \log^3 n$ 
$|W| = \phi \frac{d}{2\dispereps\gamma}$ 
for some sufficiently large constant $\phi>1$, 
$0<\dispereps<1/2$ is an arbitrary chosen constant, $d,\delta$ depend on the construction of disperser (see the construction comments later on), and $\ell=\frac{\delta}{2\dispereps\gamma}$ (recall our notation $\gamma=\frac{\threshdiff}{6}$).
Formally, a disperser $G$ is a bipartite graph satisfying the following criteria:
\begin{description}
    \item[\sf Left-degree:]
%for some $\ep>0$, 
$G$ has left-degree $d$ (i.e., each vertex in $V$ has $d$ neighbors in $W$), 
\item[\sf Right-set:]
$|W| = \dpj{\phi}\ell d/\delta$,
%$|W|\ge \phi\cdot \ell d/\delta$ for a sufficiently large constant $\phi>1$, 
%some entropy loss $\delta$,
%$|W|=\Theta(\ell d/\delta)$, %some entropy loss $\delta$,
%and satisfies the following dispersion condition:
%More precisely, it satisfies the following two conditions:
%A bipartite graph~$H=(V,W,E)$, with set $V$ of inputs
%and set $W$ of outputs and set~$E$ of edges, 
%is a \textit{$(\ell,d,\ep)$-disperser} 
%if it has the following two properties:
%\begin{quote}
%\begin{quote}
%\begin{itemize}
\item[\sf Dispersion:]
for every $L\subseteq V$ such that $|L|\ge \ell$,
the set $\Gamma_G(L)$ of neighbors of $L$ in graph $G$ is of size at least $(1-\dispereps)|W|$.
\end{description}
%%%%%%$(\phi d/\gamma \cdot 2\log^3 n, \log^3 n, \varepsilon)$-disperser 
%$(\frac{\delta}{2\dispereps\gamma}, d, \dispereps)$-disperser graph $G = (V,W,E)$, where $|V| = n$ and 
%%%%%%%$|W| \ge \frac{c}{\gamma} \cdot \log^3 n$ 
%$|W| = \phi \frac{d}{2\dispereps\gamma}$ for some sufficiently large constant $\phi>1$, which can be explicitly constructed as in~\cite{TUZ} Theorem~\ref{t:TUZ}. 
%%%%%%$|W| = \Theta(1/\gamma \cdot \log^3 n)$. 
%%%%%%Assume that $|W| \geq 2/\gamma \cdot \log^3 n$ [[TODO is this always possible?]] for sufficiently large $n$.
\cite{TUZ} showed a construction of a disperser with construction parameters: left-degree $d=\polylog n$ and 
%expansion 
entropy
loss $\delta=O(\log^3 n)$, such that for each $v \in V$
%, its 
neighborhood 
$\Gamma_G(v)$ 
%$\Gamma_G(x)$ 
can be enlisted in time 
$\polylog n$.~They also mentioned 
existence of
%that there are 
dispersers~with~$d,\delta=$~$O(\log\;{n})$.

Each element $x\in N$ is associated with a unique vertex $v_x \in V$ and each node $w_g \in W$ is associated with a group counter $g$, where the group is the set $\Gamma_G(v_x)\subseteq V$ of neighbors of $w_g$ in $G$. The set of group counters $\cG(x)$ of an element $x\in N$ is the set of group counters associated with neighbors of $v_x$ in graph $G$, i.e., $\cG(x)=\{g: w_g\in \Gamma_G(v_x)\}$. 
We will be using $v_x$ and $x$ interchangeably, whenever it does not raise any confusion; similarly, $w_g$ and $g$.

\paragraph{{\em Individual entries.}}
To account for operations on elements that need to be counted precisely, we introduce a new structure $\cC$ of \emph{individual entries} with supporting procedures. The entries will be kept for two types of elements: candidates (potential hot elements with large number of insertions minus deletions) and recently modified elements (elements for which there exists an operation insert or delete at most  $\lceil 2/\gamma \rceil$ steps ago). The reason why we need to keep individual entries for recently modified elements is that we do not process an incoming operation immediately. Instead, 
%in our algorithm upon arrival of an operation we only pre-process it, 
upon arrival of an operation our agent %only pre-processes it,
inserts it to an auxiliary queue $\mathcal{Q}$ and updates group counters and individual entries (at most) $\lceil 2/\gamma \rceil$ steps later. All these happen in the local memory of the agent (of limited capacity) and does not cause the agent to go backwards the stream nor looking ahead.
%
%will use two such structures: $\mathcal{C}^{(c)}$ for elements that are candidates for being hot elements and $\mathcal{C}^{(q)}$ for elements that are inserted or deleted in recent operations in the stream. Each individual entry contains five components:
%The elements  are counted in a separate data structure of . 
Individual entry of an element $x$ consists~of:
\vspace*{-0.6ex}
\begin{itemize}[itemsep=0ex]
\item \emph{key} $x$ of the element $x\in N$;
\item \emph{candidate counter} $c(x)$, which is incremented/decremented by $1$ upon handling of each insertion/deletion of element~$x$;
%\item \emph{offset estimate} $c_0(x)$, which stores the initial value of the individual counter of element $x$ (i.e., the very first value of the individual counter of $x$ in the newly created entry, and it does not change); 
\item \emph{recent counter} $r(x)$, which stores the number of insertions minus the number of deletions of element $x$ in recent operations in the stream;
\item \emph{number of operations} $\ops{x}$, which stores the number of recent operations on element $x$.
\end{itemize}

%When checking the existence of a counter of element $x$, the structure will return both the current value $c(x)$ and the recent counter $r(x)$ and the number of recent operations $o(x)$. 

\vspace*{-0.6ex}
In each step, 
%of our algorithm, we 
our agent receives a single operation from the stream. 
%This means that in $t$ steps of the algorithm, it receives $t$ operations. 
Note that after the algorithm processes an operation, it cannot go back to it (i.e., its algorithm only makes a single online pass over the stream). 
In the processing, the following procedures are used on datastructure $\cC$.
\vspace*{-0.6ex}
\begin{description}[leftmargin=0.5em,itemsep=0ex]
\item[$\mathsf{check}(x)$:] it checks if there is an entry with key $x$. If there is, it returns $\langle c(x), r(x), \ops{x} \rangle$, otherwise it returns $null$; it takes time $O(\log |\cC|)$;
\item[$\mathsf{add}(x)$:] it adds an entry element with key $x$ and initial values of all counters $c(x)$, $r(x)$, $\ops{x}$ equal to $0$; it takes time $O(\log |\cC|)$; 
\item[$\mathsf{size}()$:] it returns the number of entries in the structure; it is a constant time operation;
%\item $\mathsf{remove}(x)$: it removes an entry with key $x$, provided it exists; \dk{[[DO WYRZUCENIA?]]}
\item[$\mathsf{apply\_recent}(\mathsf{op}, x)$:] it applies operation $\mathsf{op}$ on the recent counter $r(x)$ of element with key $x$; it checks if there is an entry with key $x$ in the balanced tree $\cT$; if there is no such entry, it creates it; then it increments $\ops{x}$ and increments or decrements (depending on the type of $\mathsf{op}$) the recent counter $r(x)$; it takes time $O(\log |\cC|)$;  
\item[$\mathsf{rollback\_recent}(\mathsf{op} ,x)$:] it performs rollback of operation $\mathsf{op}$ on an entry with key $x$; it accesses entry with key $x$; decrements $\ops{x}$, decrements (if $\mathsf{op} = \mathsf{insert}$) or increments (if $\mathsf{op} = \mathsf{delete}$) $r(x)$; it takes time $O(\log |\cC|)$;  
\item[$\mathsf{apply\_candidate}(\mathsf{op}, x)$:] it applies operation $\mathsf{op}$ on the candidate counter of entry with key $x$; it finds a copy of an entry with key $x$ in the balanced tree and increments (if $\mathsf{op} = \mathsf{insert}$) or decrements (if $\mathsf{op} = \mathsf{delete}$) its counter $c(x)$ by $1$;  it takes time $O(\log |\cC|)$; 
\item[$\mathsf{remove\_if\_small}(x,s)$:] it removes element key $x$ if the minimum group counter of $x$ is below threshold $s$ and if the number of recent operations equals to zero (i.e., $\lambda(x) = 0$); it takes time $O(\polylog n)$ (because this is the time to access all the group counters to which $v$ belongs); 
%, and returns its key $z$ and the pair $\langle c(z), c_0(z) \rangle$;
\item[$\mathsf{get\_larger\_than}(s)$:] it returns a list of keys for which the corresponding candidate counter plus recent counter exceeds $s$; it accesses the entries in order (the structure is sorted by $c(x) + r(x)$) and returns the elements as long as the value is greater than $s$. It takes time proportional to the number of returned elements.
\end{description}

\paragraph{Implementation of datastructure $\mathcal{C}$.} Implementation of datastructure $\mathcal{C}$ with time complexities of individual operations, as claimed in the previous paragraph, can be achieved using standard datastructures. We use a balanced Binary Search Tree~$T_1$ (e.g., Red-Black Tree, c.f.,~\cite{Aho1983}), with keys being the identifiers of all the elements for which an individual entry exists and entries being tuples of a type $\langle r(x), c(x), \lambda(x), p_x\rangle$. This ensures that searching and returning the individual entry of an element (if it exists) takes logarithmic time. Each entry of $T_1$ has a pointer $p_x$ to an entry in a second Balanced Binary Search Tree $T_2$ with keys being the values of $r(x) + c(x)$ and entries being tuples $\langle x, q_x\rangle$, where $q_x$ points towards the entry of element $x$ in tree $T_1$. Clearly, any update of counters $r(x), c(x), \lambda(x)$ for some element $x$ requires logarithmic time operation to find the entry in $T_1$ and then logarithmic time to modify the key of 
a corresponding
%some 
entry in $T_2$. Inserting and deleting elements  also requires logarithmic time. Listing the elements (operation $\mathsf{get\_larger\_than}(s)$) quickly is feasible since the inorder traversal of tree $T_2$ returns the elements in order of decreasing $r(x) + c(x)$. The time of the operation is proportional to the number of returned~elements, multiplied by $O(\log\; n)$.

\subsection{Main Algorithm}

\paragraph{Intuitions.} There is a following interplay between the two data structures (group counters and individual entries) in the agent's algorithm. The group counters determine which elements should be included in the individual entries as candidates for hot elements. The individual entries keep track of the operations on the candidates for hot elements and determine which elements should be returned as hot.

%\paragraph{Notation.} In the subsequent description and analysis of our algorithm we will use notation $\gamma = \threshdiff/6$. By $t$ we will denote the current step number (i.e., the number of operations of the stream that have already been handled by the algorithm).

%\vspace*{-1mm}
\paragraph{Processing of an operation.}
In our algorithm, each element $x \in N$ has an associated set of group counters $\mathcal{G}(x)$ (note that each group counter is shared by multiple elements). Processing of operation $\mathsf{op}$($x$) (where $\mathsf{op}$ is either $\textsf{insert}$ or $\textsf{delete}$) involves updating (incrementing or decrementing, resp.) each of the counters $\mathcal{G}(x)$. If all the group counters $\mathcal{G}(x)$ are above a threshold $\gamma \cdot t$, it indicates that the element may be a candidate for being a hot element. In this case, the agent checks the structure of individual entries $\mathcal{C}$ and updates the candidate counter $c(x)$ in the entry of $x$, or creates it if it does not exist. In our algorithm we define the \emph{candidate} as an element with at least $\gamma \cdot t$ net occurrences (insertions minus deletions) in the first $t$ operations of the stream.

%\vspace*{-1mm}
\paragraph{Group counters.}
A group counter $g$ is incremented when any element 
%out of some set of elements 
from the group
(recall that the group
%set 
is determined by the topology of the used disperser $G$) is inserted in the stream. This means that 
%potentially 
some element $x$ might have all its group counters $\mathcal{G}(x)$ above $\gamma \cdot t$ while its 
%number of number of 
net occurrence $n_t(x)$ 
%might 
could
be below $\gamma \cdot t$ (because insertions of other elements have caused the counters in $\mathcal{G}(x)$ to exceed the $\gamma \cdot t$ threshold). Hence, some 'false positive' candidates might be included in~$\mathcal{C}$. In our analysis we will bound the number of such false positives. We ensure this by proving that in any set of size 
$\Theta(\delta/\gamma)$,
%$\Omega(\log^3n  / \gamma)$, 
some element will have a group counter with value at most $\gamma \cdot t$. (Recall $\delta$ is the disperser's entropy loss.)
\setlength{\floatsep}{4pt}
\setlength{\textfloatsep}{1pt}
\begin{algorithm}[t]
$t \leftarrow $ current step number\;
$\gamma \leftarrow \threshdiff / 6, \; \tau \leftarrow \lceil 1 / \gamma \rceil$\;
%$s \leftarrow \left $\;
\tcp{Phase 1}
%$size \leftarrow $\;
Divide elements included in $\mathcal{C}$ into $\tau$ chunks $E_1,\dots,E_{\tau}$ each of (at most) $\lceil \mathcal{C}.\mathsf{size}() / \tau \rceil$ elements\;
%$E_i \leftarrow $ the $i$-th chunk of (at most) $\lceil size / \tau \rceil$ elements from $\mathcal{C}$\;
\For{$i\leftarrow 1$ \KwTo $\tau$\label{l:phase1_beg}}{
Receive the next operation $\textsf{op}(x)$\;
$\mathcal{Q}.\texttt{enqueue}(\textsf{op}(x))$\label{l:enq1}\;
$\mathcal{C}.\mathsf{apply\_recent}(\mathsf{op},\;x)$\;
%$E_i \leftarrow $ the $i$-th chunk of (at most) $\lceil size / \tau \rceil$ elements from $\mathcal{C}$\;
\ForEach{$v \in E_i$}{
$\mathcal{C}.\mathsf{remove\_if\_small}(v,\left\lceil\gamma t\right \rceil)$\label{l:phase1_end}\;
}
}
\tcp{Phase 2}
\For{$i\leftarrow 1$ \KwTo $\tau$\label{l:phase2_beg}}{
Receive the next operation $\textsf{op}(x)$\;
$\mathcal{Q}.\texttt{enqueue}(\textsf{op}(x))$\label{l:enq2}\;
$\mathcal{C}.\mathsf{apply\_recent}(\mathsf{op},x)$\;
\tcp{Process two items from the queue}
$Process(\mathcal{Q.\texttt{dequeue}()},\;\mathcal{\mathcal{C}},\;\gamma,\;t + 2(i-1)$)\; 
$Process(\mathcal{Q.\texttt{dequeue}()},\;\mathcal{\mathcal{C}},\;\gamma,\;t + 2(i-1) + 1$)\label{l:phase2_end}\;
\vspace*{-1mm}
}
\caption{Handling $2 \lceil \frac{1}{\gamma}\rceil$ consecutive operations.}
\label{alg:main}
\end{algorithm}
\begin{algorithm}[t]
\caption{Extracting hot elements at step $t$.\label{alg:2}}
\KwRet{$\mathcal{C}.\mathsf{get\_larger\_than}((\secondthresh - \threshdiff)t)$}\;
\end{algorithm}

\begin{procedure}[ht]

\caption{Process($\mathsf{op}$($x$), $\mathcal{C}$,$\gamma$, $t$)}
%\caption{Extracting hot elements at step $t$.\label{alg:2}}
\tcp{Processing a single operation}
%\tcp{A subprocedure to process a single operation.}
Compute set $\cG(x)$ of group counters of $x$\label{l:group-counters-x}\;
\ForEach{$g \in \cG(x)$}{
    \lIf{\textsf{op} = \textsf{insert}}{$g \leftarrow g +1$ \label{l:group-increase-x}}
	\lElse{$g \leftarrow g  - 1$\label{l:group-decrease-x}}
}
$\mathcal{C}.\mathsf{rollback\_recent}(\mathsf{op},\;x)$\;
\If{$\mathcal{C}.\mathsf{check}(x)$\label{l:check}}
{
	$\mathcal{C}.\mathsf{apply\_candidate}(\textsf{op},\;x)$\;
%	\lIf{\textsf{op} = \textsf{insert}}{$\mathcal{C}.\mathsf{increment}(x)$\label{l:insert}}
%	\lElse{$\mathcal{C}.\mathsf{decrement}(x)$\label{l:delete}}
}
\Else{
	$g_{\min} \leftarrow \min\{g : g \in G(x)\}$\;
	\If{$g_{\min}\geq \lceil \gamma t \rceil$\label{l:qualify}}
	{
		$\mathcal{C}.\mathsf{add}(x)$\label{l:add}\;
		$\mathcal{C}.\mathsf{apply\_candidate}(\textsf{op},\;x)$\label{l:apply}\;
		
	}
}
\vspace*{-1mm}
%\caption{Handling of operation in step $t$.\label{alg:1}}
\label{proc:process}
\end{procedure}

\paragraph{Candidates.} The key property  guarantying that our algorithm is correct is the following: in every step $t$, each element with at least $\gamma \cdot t$ net occurrences has its individual entry in $\mathcal{C}$. It is nontrivial to show, as this property must hold in every step, and the set of such elements might change over time. We have to make sure that when an element exceeds the threshold, an individual entry for this element must be created immediately. We ensure this by $(1)$ keeping individual entries for all elements for which there is even a single recent operation, and $(2)$ inserting an element into $\mathcal{C}$ when each of its group counters exceeds $\gamma \cdot t$. With these properties, we know that each hot element is at any time in $\mathcal{C}$ and, moreover, at the moment such an element is inserted into $\mathcal{C}$ it is just crossing the $\gamma \cdot t$ threshold. Because the occurrences are counted precisely after an element is added to $\mathcal{C}$, we miss only $\gamma \cdot t_x$ net occurrences of an element $x$ inserted into $\mathcal{C}$ at step $t_x$. Since $\gamma < \threshdiff$ and $t_x \leq t$, each hot element has at least $(\secondthresh - \threshdiff) \cdot t$ net occurrences while having an individual entry. This allows our algorithm to correctly find all the hot elements.
\iffalse
\paragraph{Listing the hot elements.}
Returning of the hot elements is straightforward, as it only involves returning all the elements from $\mathcal{C}$ for which the sum of counters $r(x) +c(x)$ exceed a threshold $(\secondthresh - \threshdiff) \cdot t$. We assure that we return all hot elements, however we may also return some other elements. However, the additional returned elements have surely at least $(\secondthresh - \threshdiff) \cdot t$ net occurrences.
\fi
\paragraph{Pseudo-codes.}
The algorithm is described on three pseudo-codes: one handling 
%a window of 
$2\lceil 1/\gamma\rceil$ consecutive operations in the stream, c.f., a pseudo-code in Algorithm~\ref{alg:main}, a sub procedure that processes a single operation, c.f., procedure Process, and the final pseudo-code enlisting hot elements in the processed stream, c.f., a pseudo-code in Algorithm~\ref{alg:2}. Note that a request to return hot elements can arrive at any step (also while
%in the middle of 
handling 
%a window of 
$2\lceil 1/\gamma\rceil$ consecutive operations). In such case, the execution of Algorithm~\ref{alg:main} is paused
and Algorithm~\ref{alg:2} is immediately executed. 

The latter is a single execution of procedure $\mathsf{get\_larger\_than}(s)$ for 
%parameter 
$s=(\secondthresh - \threshdiff) t$, where $t$ is the current position~in~the~stream.

The former, Algorithm~\ref{alg:main}, works as follows. In a single iteration of the algorithm, %it the agent
processes 
%a window of 
$2\lceil 1/\gamma\rceil$ consecutive operations, online one-by-one. 
%Each 
This part of the stream, also called
a window (recall though the agent does not process whole window at once, but online one-by-one), is further divided into two sub-windows of $\lceil 1/\gamma\rceil$ operations each. In the first sub-window the agent performs Phase 1, during which it executes the cleanup of $\mathcal{C}$ (see Algorithm~\ref{alg:main} lines \ref{l:phase1_beg}-\ref{l:phase1_end}). 
%Note that 
During these steps, the new operations are pre-processed (using $\mathsf{apply\_recent}$ method of $\mathcal{C}$) and inserted into the queue $\mathcal{Q}$. In each step of processing the second sub-window (Phase 2), the agent finishes processing of some pair of operations from the queue (see Algorithm~\ref{alg:main} lines \ref{l:phase2_beg}-\ref{l:phase2_end}). %With such approach, 
Hence, the length of the queue decreases by $1$ in each step of Phase~2, and thus after the last step 
%of the subwindow, 
the queue is empty. Note that our algorithm, even though it processes operations 
in windows, is strictly online, because in each step the agent receives 
a new operation from the stream and each received operation is 
immediately handled.

%%%%%%%%%%%%  END  REMOVE  %%%%%%%%%%%%%%

\section{Analysis of the Algorithm}
\label{s:analysis}

 %Consider $g$ a group counter for some set of elements $S\subset N$. 
 %At any step of the algorithm we can consider the value of the group counters with added values for elements, for which we have the individual counters $\bar{g} = g + \sum_{s\in S} (c(s) - c_0(s))$ (for simplicity, we assume that if for some element $s' \in S$ there is no individual counter, then $c(s') - c_0(s') = 0$).

In the analysis, for the sake of clarity, we sometimes add lower index $t$ to some model variables to indicate that the values of these variables are taken just after processing $t$ operations from the stream, e.g., $\cG_t(x)$ denotes the set of all group counters of element $x$ after processing the first $t$ operations.

The first lemma shows that there is a limited number of elements $x\in N$, namely 
$\frac{\delta}{2\dispereps\gamma}$,
%$2 / \gamma \cdot \log^3 n$, 
having all its group counter at least $\lceil \gamma t\rceil$. The second lemma shows that the number of recent operations stored in the recent counters is at most $\lceil 1/\gamma \rceil +1$.
%It implies that if the number of individual entries is at least 
%$\frac{\delta}{\dispereps\gamma}$,
%$2 / \gamma \cdot \log^3 n$, 
%then at least half of these element must have a group counter smaller than $t\gamma$. 
Thus procedure $\mathsf{remove\_if\_small}$ executed on all the elements from $\cC$ removes all but at most $\frac{\delta}{2\dispereps\gamma} + \lceil 1/\gamma \rceil +1$ entries.
%will remove half of all the elements from $\mathcal{C}$. 
When an operation arrives we first \emph{pre-process} it by inserting into queue $\mathcal{Q}$ (and upon insertion, this operation is accounted for in a recent counter). Some number of steps later we 
%\emph{process}
\emph{finish processing} the operation, %when we execute the sub-procedure
by executing procedure
$Process$ with this operation as the first argument. 

\begin{lemma} 
\label{lem:1}
After processing first $t$ stream operations by Algorithm~\ref{alg:main}, for any set $S \subseteq N$, where $|S| = \frac{\delta}{2\dispereps\gamma}$,
%\geq 2 / \gamma \cdot \log^3 n$, 
% elements, 
there is an element $s \in S$, such that 
$g < \gamma t$ for some of its group counters $g \in \cG_t(s)$.
%$g_t(s) < \gamma t$ for some of its group counters $g_t(s) \in \cG_t(s)$.
\end{lemma}

\vspace*{-1ex}
\begin{proof}
First observe, that since in upon processing an operation, we increment at most 
%$d = \log^3 n$ 
$d=\polylog n$ group
counters 
%$\bar{g}$, 
(recall that $d$ is the left-degree of the disperser $G$ used for the construction, c.f.,~\cite{TUZ}),
then after processing $t$ operations from the stream the sum of the group counters is at most 
%$t \log^3 n$. 
$t d$.
Since there are 
%$2/\gamma \cdot \log^3 n$ 
%$|W|=cd/\gamma$
\dpj{$|W|=\frac{\phi d}{2\dispereps\gamma}$}
group counters (\dpj{where a constant $\phi > 1$}), associated 1-1 with the right vertices of disperser $G$, then the average value of a group counter is at most 
%$\gamma t/2$. 
$td/|W|$. 
At least 
$2\dispereps$-fraction
%half 
of the counters must be below 
$1/(2\dispereps)$
%two 
times the average; thus, there are at least 
%$|W| / 2$ 
$2\dispereps |W|$
group counters with values smaller than 
$td/(2\dispereps |W|)\le\gamma t$, where $\dispereps <1/2$ is a parameter of the disperser $G$ and $W$ of size $|W|\ge\frac{d}{2\dispereps\gamma}$ is the number of its right-side vertices.  
If we take any set $S\subseteq N$ of size at least 
%$1 / \gamma \cdot 2 \log^3 n$, 
%elements from $W$, 
$\frac{\delta}{2\dispereps\gamma}$,
the set $N_G(S)$ of neighbors of $S$ in graph $G$ is of size at least $(1 - \dispereps)|W|$, by dispersion.
%> |W| / 2$. 
Hence 
at least  $2\dispereps |W| + (1 - \dispereps)|W| - |W|=\dispereps |W|$
%some of the 
counters associated with some elements in $S$ must be below $\gamma t$. 
\end{proof}

\begin{lemma}
\label{lem:recent1}
At any step of the algorithm, the total number of operation stored in recent counters is at most $\tau +1$.
%(i.e., $\sum_{v \in N} o(v) \leq \tau + 1$).
\end{lemma}

\vspace*{-1ex}
\begin{proof}
It is easy to see that the size of queue $\mathcal{Q}$ in Algorithm~\ref{alg:main} is always at most $\tau + 1$. It is because it is empty at the beginning of the algorithm and then we add a single operation to it in each iteration of for-loop in lines \ref{l:phase1_beg}-\ref{l:phase1_end} of Algorithm~\ref{alg:main} and then in each iteration of the for-loop in lines \ref{l:phase2_beg}-\ref{l:phase2_end} of Algorithm~\ref{alg:main} we first add a single operation and then remove two. Since both for-loops have $\tau$ iterations each, the maximum size of the queue is after line \ref{l:enq2} during the first iteration of the second for-loop and equals to $\tau + 1$. Hence the total number of operations stored in the recent counters is also at most $\tau +1$. It remains to observe that each operation in the queue contributes exactly one to the sum $\sum_{v \in N} \lambda(v)$.
\end{proof}

Using Lemmas~\ref{lem:1} and~\ref{lem:recent1} we can show that the size of the datastructure $\mathcal{C}$ is always $O(\delta/\threshdiff)$.
%$O(\log^3 n/\threshdiff)$.

\begin{lemma}
\begin{enumerate}[itemsep=0ex]
    \item After each Phase 1, the number of individual entries in $\mathcal{C}$ is at most $\frac{\delta}{2\dispereps \gamma} +\lceil 1/\gamma\rceil + 1$;
    \item After each Phase 2, the number of individual entries in $\mathcal{C}$ is at most $\frac{\delta}{2\dispereps \gamma} + 3 \lceil 1/\gamma\rceil + 1$.
\end{enumerate}
\end{lemma}

\vspace*{-1ex}
\begin{proof}
To prove (1), note that, we do not update the group counters and do not add any new individual entry during this phase. Moreover, by Lemmas~\ref{lem:1} and~\ref{lem:recent1}, at the beginning of Phase 1, there are at most $\frac{\delta}{2\dispereps \gamma} +\lceil 1/\gamma\rceil + 1$ elements that will not be removed by procedure $\mathsf{remove\_if\_small}$. To prove (2), observe that we add at most $2\tau$ new individual entries during Phase 2. 
\end{proof}

%At any step $t$ and any element $x \in N$, we can consider the following value. We take the set $G_t(x)$ of group counters for $x$ at step $t$. Let $\bar{n}_t(x) = min_{g_t\in G_t(x)} \bar{g_t}$. The following lemma shows that from our counters we can calculate an upper bound of the number of occurrences of any element.

\noindent
%The next lemma relates 
Next we relate
the number of net occurrences $n_t(x)$ of an element $x\in N$ in the stream by position $t$, to the counters $c(x)$ and $r(x)$.

\begin{lemma}
\label{lem:2}
After first $t$ steps of Algorithm~\ref{alg:main}:
%In any step $t$ of the algorithm:
\begin{enumerate}[itemsep=0ex]
\item for every element $x \in N$, if $n_t(x) \geq \lceil \gamma t \rceil$, then $x\in \mathcal{C}_t$ ,
\item for every element $x \in \mathcal{C}_t$ we have 
$n_t(x) -  \lceil \gamma t \rceil \leq c_t(x) + r_t(x) \leq n_t(x)$.
%\item for every element $x$ we have $g_t + r(x)\geq n_t(x)$ for each group counter $g_t \in \cG_t(x)$.
%\item for every element 
\end{enumerate}
\end{lemma}

\vspace*{-1ex}
\begin{proof}
To show $(1)$, we need to show, that in the algorithm after processing $t'$ operations all elements with at least $\lceil t' \gamma \rceil$ occurrences are in $\mathcal{C}$. To prove this we need to show that we never remove from $\cC$ an element with at least $\lceil t'\gamma \rceil$ net occurrences and that we always add to $\cC$ an element when its number of net occurrences becomes equal to or larger than $\lceil t'\gamma \rceil$. We prove this fact by induction. Assume that it holds for all phases up to some step $t'$. We then consider the following Phase $1$ and Phase~$2$. Note that the queue $\mathcal{Q}$ is empty at the beginning of each Phase~$1$. It is because each Phase $1$ adds $\tau$ operations to it and each Phase $2$ removes $\tau$ operations from the queue. Note that, when the queue is empty, each group counter contains the total number of insertions minus number of deletions of all the elements from the group (because we processed all the operations up to this step). This means that any element removed from $\mathcal{C}$ by operation $\mathsf{remove\_if\_small}$ has some group counter at most $\lceil t \gamma \rceil$, hence it surely has at most $\lceil t \gamma \rceil$ occurrences in the first $t$ operations in the stream. Additionally, we keep in $\mathcal{C}$ individual entry for each element with at least one (recent) operation in the interval $[t,t+\tau]$. Hence after Phase $1$ we have in $\mathcal{C}$ all the elements that have at least $\lceil t \gamma \rceil$ occurrences in the first $t$ operation or at least one operation in the interval $[t,t+\tau]$. Hence after Phase 1 we have that $\mathcal{C}$ contains all the elements that have at least $\lceil(t+\tau)\gamma \rceil$ occurrences. %In Phase $2$, we may remove an element only in procedure $\mathsf{rollback\_recent}$ upon handling of some $t'$-th operation (for some $t +2\tau > t' > t$). This happens if the $o(x)$ counter of some element $x$ drops to $0$ and some of its group counters is below $\lceil t' \gamma \rceil$. But this means that this element surely has at most $\lceil t' \gamma \rceil$ occurrences in the first $t'$ operations of the stream. 
 In Phase $2$ we add elements to $\mathcal{C}$ only upon handling of some $t'$-th operation $\mathsf{op}(x)$ if each group counter of $x$ equals to at least $\lceil t'\gamma\rceil$. Hence, after handling $t'$-th operation in Phase 2, the claim holds. This 
 finishes
 %completes 
 the inductive~proof. 

    To show $(2)$, we first observe that if an element $x \in \mathcal{C}$, then upon processing an opration on $x$, this operation is immediately accounted for in counter $r(x)$. Hence $r(x)$ is incremented/decremented upon each insertion/deletion of element $x$ as long as $x$ belongs to $\mathcal{C}$. Note that upon handling of an operation, the sum $r(x) + c(x)$ does not change. Hence $r_t(x) + c_t(x)$ stores the number insertions minus the number of deletions of element $x$ since the step when $x$ was inserted into $\mathcal{C}$. This means that $c_t(x) + r_t(x) \leq n_t(x)$. To show that left side of $(2)$, consider the largest time step $t' < t$, such that at step $t'$, $x\notin \mathcal{C}$. We know from $(1)$, that $n_{t'}(x) < \lceil\gamma t'\rceil$. We also know that as long as  $x$ belongs to $\mathcal{C}$, $c_t(x) + r_t(x)$ counts all the occurences of $x$, hence $c_t(x) + r_t(x) = n_t(x) - n_{t'+1}(x) \geq n_t(x) - \lceil \gamma t' \rceil$.
%
    %that if $n_t(x) \leq \lceil t\gamma \rceil$ then the claim is tivially true. 
    %TODO finish this
%
%We prove this lemma by induction.  The lemma clearly holds before processing any operation. Consider any operation on element $x$.  
% Consider any operation on element $x$. Clearly we increment or decrement each group counter of element $x$, hence 2. holds for all elements.
% 
% If $x$ belongs to $\mathcal{C}$, then by the definition of the algorithm (lines 7,8) after processing this operation, the lemma will hold. 
% If $x$ is added to $\mathcal{C}$ in this step, then the lemma also holds for $x$ because, the initial value of the individual counter equals to the minimum value of the group counters from the set $G(x)$ (after we increment/decrement each counter from this set). 
%
% If we remove an element $z$ from $\mathcal{C}$, then consider the last step $t' <t$, when $z$ was added to $\mathcal{C}$. We have $n_t(z) = n_{t'}(z) + c(z) - c_0(z)$. By the Basic Integrity Constraint, for each group counter $g_t \in G_t(x)$ we have $g_t \geq n_{t'}(z)$, hence after adding $ c(z) - c_0(z)$ (line 16) to each of the counters the lemma holds for $z$. 
\end{proof}

\begin{theorem}
Algorithms~\ref{alg:main} and~\ref{alg:2} correctly solve the problem of deterministically finding $(\secondthresh,\;\threshdiff)$-hot elements using $O(\log^3 n/ \threshdiff)$ memory,  
$O(\polylog n)$  worst-case time for processing a single operation in the stream, and $O(\log\; n/\secondthresh)$ time for enlisting  the $(\secondthresh,\;\threshdiff)$-hot elements. 
%(there are $O(1\secondthresh)$ of them).
%[[Pamietajmy o listowaniu sasiedztwa w disperserach]]}
%$O(\log n)$ time per operation. 
\end{theorem}

\vspace*{-1ex}
\begin{proof}
To show that the main algorithm, consisting of Algorithms~\ref{alg:main} and~\ref{alg:2}, is correct we need to show two facts:
\vspace*{-0.3ex}
\begin{enumerate}[itemsep=0ex]
\item it %the algorithm 
returns all elements with at least $t \secondthresh$ occurrences,
\item it 
%the algorithm 
returns no element with less than $t (\secondthresh - \threshdiff)$ occurrences.
\end{enumerate}
\vspace*{-0.5ex}
 The second fact follows directly from Lemma~\ref{lem:2}(2), because if $t(\secondthresh - \threshdiff) > n_t(x)$, then also $t(\secondthresh - \threshdiff) >c(x) + r(x)$, and such element is not returned by procedure $\textsf{get\_larger\_than}$.

The first part follows from Lemma~\ref{lem:2}(1). If $t\leq 1/ \gamma$, then all the insertions minus deletions are stored in the recent counters (i.e., $r_t(x) = n_t(x)$), hence if $n_t(x) \geq t\secondthresh$, then $r(x) \geq t (\secondthresh - \threshdiff)$ and this element will be returned by procedure $\mathsf{get\_larger\_than}()$.

Otherwise, if $t> 1 / \gamma$ if we have an element $x$, with $n_t(x) > t\secondthresh = t(\secondthresh -\threshdiff) + 6\gamma t > t(\secondthresh -\threshdiff) + t\gamma +5 > t(\secondthresh -\threshdiff) = t(\secondthresh -\threshdiff) + \lceil t\gamma \rceil  $. Then by Lemma~\ref{lem:2}(1) $x\in \mathcal{C}_c$ and $c_t(x) + r_t(x) \geq n_t(x) - \lceil \gamma t\rceil \geq  t (\secondthresh -\threshdiff)$ and this element will be returned by procedure $\mathsf{get\_larger\_than}()$.

The memory complexity of our algorithm is straighforward from the definition of the data structures (the dominating part is the structure of group counters, and the overhead $\log^3 n$ comes from the value of $\delta$ in the constructive disperser). The time of enlisting the hot elements 
%in our algorithm 
is in fact: 
$O(1/(\secondthresh - \threshdiff))$ elements times $O(\log\; n)$ per element (see the definition of procedure $\mathsf{get\_larger\_than}()$), which clearly gives $O(\log n/\secondthresh)$ for $\secondthresh-\threshdiff=\Theta(\secondthresh)$; in case $1/(\secondthresh - \threshdiff) = o(1/\secondthresh)$ we can run our algorithm for finding $(\secondthresh, \; \secondthresh /2)$-hot elements rather than original parameters $\secondthresh,\threshdiff$, which clearly returns only valid $(\secondthresh,\; \threshdiff)$-hot elements. Polylogarithmic time per operation follows from the polylogarithmic time of enlisting group counters of a single element (using disperser) and a constant number of calls to logarithmic-time procedures on the individual entries structure.
%
%To see that the amortized time is $O(\polylog n)$ observe that procedure $\mathsf{remove\_small}$ removes at least $\delta / (2\dispereps\gamma)$ elements hence can be executed at most once every $\delta / (2\dispereps\gamma)$ operations. 
\end{proof}
%Upon arrival of operation $\textsf{op}(x)$:
%\begin{enumerate}
%\item 
%Check if there is an individual counter for item $x$, by checking the search tree (we show it is in $O(\log k + \log\log n ???)$ time): if yes, read this counter as $c_t(x)$, otherwise set $c_t(x)\gets null$
%
%\item
%If $c_t(x) \ne null$
%%\gamma t$ 
%then increase/decrease the individual counter of $x$ by $1$, respectively for insertions/deletions, in the search tree and update also the heap (we show it is in $O(\log k + \log\log n ???)$ time)\footnote{%
%Instead of update by $1$, we may use multiplicities.}
%
%\item
%If $c_t(x) = null$ then  
%
%\begin{enumerate}
%\item identify the set $G_t(x)$ of group counters for $x$. This can be done in time $O(poly\log n)$ (using the properties of the disperser).
%\item increase/decrease all group counters of $x$ by $1$, respectively for insertions/deletions.
%\item while increasing, calculate the minimum value over all group counters of $G_t(x)$ (after the increase). If this value exceeds $t \cdot \gamma$ then initiate the procedure of allocating an individual counter for element $x$:
%\begin{itemize}
%\item Check if all individual counters are allocated. 
%\item If not allocate a free counter for $x$.
%\item Otherwise find the smallest element in the heap (time $O(1)$), remove its individual counter and allocate a counter for $x$. Initial value of element $x$ in the heap is $t \gamma$. 
%\end{itemize}
%\end{enumerate}
%
%\end{enumerate} 
\vspace*{-3mm}
\section{Lower Bound on 
%Agent's
Memory}
\label{sec:lower}

\begin{theorem}
Any deterministic algorithm finding $(\secondthresh,\threshdiff)$-hot elements requires memory $\Omega\left(\frac{1}{\threshdiff}\right)$, even for insertions.
%if there are only insertions in the stream.
\end{theorem}

\vspace*{-1ex}
\begin{proof}[Proof Sketch]
Let $\alpha =  \lceil\frac{1}{\secondthresh-\threshdiff}\rceil - 1$.
W.l.o.g. we may 
%We first assume 
$\alpha \in Z$.\footnote{%
If $\alpha$ is not an integer, we could first, instead of the original $\secondthresh,\threshdiff,\alpha$, prove the result for $\alpha'=\lfloor\frac{1}{\secondthresh-\threshdiff}\rfloor - 1$ and  $\secondthresh'>\threshdiff'>0$ such that: $\secondthresh'=\secondthresh$ and $\lfloor\frac{1}{\secondthresh-\threshdiff}\rfloor = \frac{1}{\secondthresh'-\threshdiff'}$. Note that $\alpha'\ge 1$ is an integer. The result for $\secondthresh,\alpha,\threshdiff$ will be asymptotically the same, as $\secondthresh=\secondthresh'$, $\threshdiff=\Theta(\threshdiff')$. It will also return $(\secondthresh,\threshdiff)$-hot elements correctly, as $\secondthresh-\threshdiff \le \secondthresh'-\threshdiff' \le \secondthresh'= \secondthresh$.
\remove{%%%%%%%  START  REMOVE  %%%%%
\dk{If $\alpha$ is not an integer, we could first prove the result for $\secondthresh'=\lfloor\secondthresh\rfloor$, $\threshdiff'=$ and $\alpha'=\lceil\frac{1}{\secondthresh'-\threshdiff'}\rceil - 1$ instead of $\secondthresh,\threshdiff,\alpha$. Note that $\secondthresh'\le \secondthresh$, thus $\alpha'\ge 1$ is an integer. The result for $\secondthresh,\alpha,\threshdiff$ will be asymptotically the same, as $\secondthresh=\Theta(\secondthresh')$, $\threshdiff=\Theta(\threshdiff')$. It will also return $(\secondthresh,\threshdiff)$-hot elements correctly, as $\secondthresh-\threshdiff \le \secondthresh'-\threshdiff' \le \secondthresh'\le \secondthresh$.}
}%%%%%%%  END  REMOVE  %%%%%%%%%%%%
}
%%%%%%%$1/\firstthresh \in \mathbb{Z}$.
%
Consider any deterministic algorithm finding $(\secondthresh,\; \threshdiff)$-hot elements.
We design a stream to enforce the lower bound %
%$\Omega\left(\frac{1+\secondthresh - \threshdiff}{\threshdiff}\right)=
$\Omega\left(\frac{1}{\threshdiff}\right)$
%$\Omega\left(\frac{1-(\secondthresh - \threshdiff)}{\threshdiff}\right)$ 
on the size of algorithm's memory. 

Let $x = \left\lfloor \frac{1- \secondthresh + \threshdiff}{\threshdiff}\right\rfloor$,
%$x = \left\lfloor \frac{1- (\secondthresh - \threshdiff)}{\secondthresh - \threshdiff}\right\rfloor$, 
and let $X$ be an arbitrary set of $x$ elements.
%Take any set $X$ of $|X| = x = \left\lfloor \frac{\left(1- \frac{1}{\lceil 1/\firstthresh\rceil}\right)}{\secondthresh - \firstthresh}\right\rfloor -1$ elements from $N$. 
Note that $x = \Theta(1/\threshdiff)$.
First part of the stream $\mathcal{S}(X)$ associated with set $X$ contains $\alpha$ insertions of each element from set $X$, starting from $\alpha$ insertions of the first element in $X$, then $\alpha$ insertions of the second element, etc. (W.l.o.g. we can assume that for the sake of analysis the elements in $X$ are ordered arbitrarily.) 
\remove{%%%%%%%  START  REMOVE  %%%%%%
We want to argue that after processing these operations, the algorithm has to remember the set $X$, 
in the sense 
% We will show 
that the algorithm has to correctly answer any query 
``{\em does element $v \in N$ belong to $X$?}''.
%whether an element $v \in N$ belongs to $X$ or not.

Obviously, the considered system does not allow to ask such queries directly (only queries regarding listing current hot elements could be executed), but we could 
realize
%implement 
this idea implicitly but formally as follows.
%The query works as follows. 
%
}%%%%%%%  END  REMOVE  %%%%%%%%%%%%%
Next, the adversary adds $x$ insertions of element $v\in N$ to the previously constructed stream and checks whether element $v$ is hot after these operations. Observe that if $v \notin X$, we have 
%\vspace*{-1ex}
%\[
$
f(v) = \frac{x}{\alpha x + x} = \frac{1}{\alpha + 1} \le \secondthresh - \threshdiff
$.
%\ .
%\]
%\vspace*{-1ex}
Otherwise, i.e., if $v\in X$, we get
\vspace*{-1ex}
\begin{align*}
f(v) 
&= 
\frac{\alpha + x}{\alpha x + x} 
= 
\frac{\alpha /x +1}{\alpha +1} 
\ge 
\frac{\frac{(1/ (\secondthresh - \threshdiff) -1)\cdot \threshdiff}{1-\secondthresh + \threshdiff} + 1}{1/ (\secondthresh - \threshdiff)} 
\\
&= 
\frac{\threshdiff/(\secondthresh - \threshdiff) +1}{1/ (\secondthresh - \threshdiff) } 
%+\frac{1 - \frac{\alpha (\secondthresh - \threshdiff)}{1- (\secondthresh - \threshdiff)}}{1/ (\secondthresh - \threshdiff)} 
%\\ 
%&\ge 
=
\frac{\secondthresh/(\secondthresh - \threshdiff)}{1/ (\secondthresh - \threshdiff) } 
%
%\secondthresh +\frac{1 - \frac{1 -(\secondthresh - \threshdiff)}{1- (\secondthresh - \threshdiff)}}{1/ (\secondthresh - \threshdiff)} 
= 
\secondthresh
\ .
\end{align*}
\remove{%%%%%%%%%%%%
\begin{align*}
f(v) &= \frac{\alpha + x}{\alpha x + x} = \frac{\alpha /x +1}{\alpha +1} 
\ge \frac{\alpha\frac{\threshdiff}{(1- (\secondthresh - \threshdiff))} + 1}{1/ (\secondthresh - \threshdiff)} 
\\
&= \frac{(1/(\secondthresh - \threshdiff) -1) \secondthresh}{(1- (\secondthresh - \threshdiff))  1/ (\secondthresh - \threshdiff) } +\frac{1 - \frac{\alpha (\secondthresh - \threshdiff)}{1- (\secondthresh - \threshdiff)}}{1/ (\secondthresh - \threshdiff)} 
\\ & \ge \secondthresh +\frac{1 - \frac{1 -(\secondthresh - \threshdiff)}{1- (\secondthresh - \threshdiff)}}{1/ (\secondthresh - \threshdiff)} 
\\
 &= \secondthresh
\ .
\end{align*}
}%%%%%%%%%%%%%%%
%\dk{Nastepny paragraf trzeba napisac formalnie, np. jako dowod nie wprost.}

\vspace*{-0.6ex}
\noindent
Since there are ${n\choose x}$ choices of set $X$, the number of bits to uniquely encode all possible sets $X$ is $\log_2 {n\choose x} = \Theta((1-(\secondthresh - \threshdiff))/\threshdiff \cdot \log\; n)$, which is $\Theta(1/\varepsilon)$ machine words. Consider an algorithm $\mathcal{A}$ for finding $(\secondthresh,\;\threshdiff)$-hot elements using less memory. Since the memory size is smaller than $\log_2 {n\choose x}$ machine words, there exist two sets $X_1 \neq X_2$, $|X_1|, |X_2| = x$, such that after processing stream $\mathcal{S}(X_1)$ the memory state is exactly the same as after processing stream $\mathcal{S}(X_2)$.  We showed that after processing $\mathcal{S}(X_1)$ all and only elements from $X_1$ can be returned by the algorithm as hot element. On the other hand, in $\mathcal{S}(X_2)$, all and only elements from $X_2$ can be returned by the algorithm as hot element. This is a contradiction with the assumption that $\mathcal{A}$ is a correct deterministic algorithm for finding  $(\secondthresh,\; \threshdiff)$-hot elements.
%
%Hence, in order to satisfy the definition of $(\secondthresh,\threshdiff)$-hot elements, the algorithm has to remember the whole set $X$. Since there are ${n\choose x}$ choices of set $X$, the minimum number of bits needed by the algorithm is $\log_2 {n\choose x} = \Theta((1-(\secondthresh - \threshdiff))/\threshdiff \cdot \log n)$, which means that the algorithm needs to use $\Theta(1/\threshdiff)$ machine words.
%
%It remains to consider the case when $\alpha$ is not an integer. Then, consider $\firstthresh'=\frac{1}{\lceil\frac{1}{\firstthresh}\rceil}$. Note that $\firstthresh'\le \firstthresh$ and $\alpha'=\frac{1}{\firstthresh'} -1$ is an integer. Thus, we could apply the above proof to $\firstthresh'$ instead of $\firstthresh$ and $\alpha'$ instead of $\alpha$.
\end{proof}
\vspace*{-3mm}
\section{Extensions, Discussion and 
Open Directions}
%Open Problems}
\label{sec:conclusions}

\paragraph{False positives, false-negatives and stochastic counterpart.}

Our approach guarantees no false-negatives, one could turn it into no false-positives by running the agent for parameters $\secondthresh+\threshdiff$ and $\threshdiff$ instead of $\secondthresh$ and $\threshdiff$, resp. 
\remove{%%%%%% START  REMOVE  %%%%%%
that all hot elements are outputted, however a limited number (i.e., $O(1/\secondthresh)$) of semi-hot elements (i.e., with frequency between $\secondthresh - \threshdiff$ and $\secondthresh$) could be returned as well. The latter could be viewed as ``false-positive'' results. If one prefers to end up with ``false-negative'' rather than ``false-positive'' elements, then our algorithm could be applied for parameters $\secondthresh + \threshdiff, \threshdiff$ instead of $\secondthresh, \threshdiff$; with the former pair of parameters our algorithm does not return elements of frequency smaller that $\secondthresh$, however it may also not return some elements with frequency in $(\secondthresh,\secondthresh +  \threshdiff)$, i.e., some hot elements with frequency just slightly above $\secondthresh$.
}%%%%%%  END  REMOVE  %%%%%%%%%%%%%%
If the stream is stochastic, our agent returns {\em all and only} elements with frequency at least $\secondthresh$ with probability corresponding to the distribution of hot items, e.g., for power-law distribution the probability of false positive/negative is polynomial in $\threshdiff$.

\vspace*{-0.05ex}
\paragraph{Graphs, hypergraphs, relationships and weights.}
Our agent could be applied if the universe $N$ contains a relationship $\cR$ on some set $\cN$, e.g., a graph or a hypergraph. 
%on the set of vertices $\cN$. 
Our agent not only monitors high frequencies of (hyper-)edges, but also vertices from set $\cN$ or sets in any other sub-relation $\cR'$ of $\cR$, with a slight performance overhead $\max_{R\in \cR} |\{R' : R'\subseteq R \ \ \& \  R'\in\cR'\}|$.
\remove{%%%%%%%%%%%  START  REMOVE  %%%%%%%%
More precisely, for any specified sub-relation $\cR'$ of $\cR$, upon analyzing an element $R\in \cR$ from the stream, the algorithm can compute all elements $R'\subseteq R$ belonging to $\cR'$ and apply the corresponding transaction (insert or delete) to all such elements $R'$.
For instance, we could monitor single elements by setting $\cR'=\{\{v\} : v\in\cN\}$, i.e., the set of all singletons.
In general, the upper bound on the time performance of processing algorithm increases by factor at most $\max_{R\in \cR} |\{R' : R'\subseteq R \ \ \& \  R'\in\cR'\}|$ and the memory increases by factor ??? 
\dk{TO BE CHECKED}
}%%%%%%%  END  REMOVE  %%%%%%%%%%%%%
%
%\paragraph{Weighted version.}
Moreover, instead of inserting or deleting elements, we could implement more complex operations of updating variables in set $N$.
%on the top of our algorithm. For example, we could implement transactions of adding or deducting numbers in the stream to specific variables.

\vspace*{-0.05ex}
\paragraph{Parallel version.} When the stream operations appear faster than a single agent can handle,
%we can handle using a single worker, 
we can employ multiple agents to process the operations. Parallelization of our algorithm is possible assuming that the agents have parallel access to queue~$\mathcal{Q}$, structure $\mathcal{C}$ and the structure of group counters. When having $\kappa$ agents, for $\kappa < \tau$, we can parallelize the loops in lines \ref{l:phase1_beg} - \ref{l:phase1_end} and \ref{l:phase2_beg} - \ref{l:phase2_end} of Algorithm~\ref{alg:main}. We handle $\kappa$ operations in each parallel step (and $\kappa\; \text{mod}\; \tau$ once every $\lceil\tau / \kappa\rceil$ parallel step) and we retain the same memory complexity of the algorithm (i.e., agents could be using a small shared memory).

\remove{%%%%%%  START  REMOVE  %%%%%%%%%

\paragraph{Tolerating more adaptiveness.}
Note that the derived correctness and performance guarantees of our algorithm hold even if some malicious adversary sets up elements of the stream on the fly, which makes our agent a perfect tool for analysis of malicious behaviors in the streams. 

\paragraph{Trading fuzziness for probability of false positives/negatives.}

\paragraph{Waiving the Basic Integrity Constraint.} Our algorithm could work under assumption that if the frequency is at most $\delta$ we do not have to execute this operation in a round.

}%%%%%%%%%  END  REMOVE  %%%%%%%%%%%

%\section{Pseudocodes of procedures and experimental comparison of algorithms}

\bibliographystyle{abbrv}
\bibliography{biblio}

\end{document}